\documentclass[letter]{aa}
\usepackage{graphicx}
\usepackage{txfonts}
\usepackage{hyperref}
\usepackage{soul}

\usepackage{xcolor}
\usepackage[normalem]{ulem}
\usepackage{orcidlink}
\usepackage{threeparttable} 

\usepackage{subfigure}

\begin{document} 

\title{The mass functions of multiple stellar populations in 47\,Tucanae}

\author{M.\,Scalco\inst{1}\fnmsep\thanks{\email{mscalco@iu.edu}}\orcidlink{0000-0001-8834-3734}
\and 
J.\,Bruce\inst{1}\orcidlink{0009-0000-9203-1653}
\and 
E.\,Vesperini\inst{1}\orcidlink{0000-0003-2742-6872}
\and 
L.\,R.\,Bedin\inst{2}\orcidlink{0000-0003-4080-6466}
}

\institute{
Department of Astronomy, Indiana University, Swain West, 727 E. 3rd Street, Bloomington, IN 47405, USA
\and
Istituto Nazionale di Astrofisica, Osservatorio Astronomico di Padova, Vicolo dell'Osservatorio 5, Padova 35122, Italy
}

\date{XXX,YYY,ZZZ}
 
\abstract
{We used a combination of \textit{James Webb} Space Telescope (JWST) and \textit{Hubble} Space Telescope (HST) observations to investigate the mass functions (MFs) of the different stellar populations in the globular cluster (GC) 47\,Tucanae (NGC\,104) over a wide stellar-mass interval. We identified the multiple populations along the lower main sequence using only JWST photometry and, in particular, colour combinations sensitive to variations in oxygen abundance. We derived the MFs of the first- and second-population (1P and 2P) stars separately and found that in the low-mass regime, the present-day MF of the 1P population is significantly steeper than that of the 2P population, which appears nearly flat. We also present the results of a dynamical model showing that this behaviour requires a different initial mass function (IMF) for the 2P and the 1P populations in the $0.1$--$0.4\,M_{\odot}$ mass range, thus providing a new fundamental constraint for the study of the star formation of multiple populations in GCs.} 


\titlerunning{Mass functions of mPOPs in 47 Tuc}
\authorrunning{M.\,Scalco et al.}
\maketitle
\section{Introduction}\label{Section1}
Globular clusters (GCs) have long been recognised as fundamental laboratories for studying stellar evolution and dynamical processes in dense stellar systems. Over the past two decades, it has become clear that GCs host multiple stellar populations (mPOPs) characterised by distinct chemical abundances along all major evolutionary sequences, from the main sequence (MS) to the red giant branch (RGB) and beyond \citep[see e.g.][for a review]{2019A&ARv..27....8G}. Understanding how these populations form and evolve and how they are affected by internal dynamical processes remains one of the central open problems in GC studies.\looseness=-4

Multiple stellar populations in GCs are commonly grouped into a first population (1P) and one or more second populations (2Ps). The 1P is generally characterised by chemical abundance patterns similar to those of halo field stars at the same metallicity, while the 2P stars show signatures of processed material such as enhancements in helium, nitrogen, and sodium and depletions in carbon and oxygen. These chemical differences manifest themselves photometrically as distinct sequences in suitable colour-magnitude and two-pseudo-colour diagrams (CMDs and TpCDs, respectively), allowing the different populations to be identified even in the absence of spectroscopy.\looseness=-4

One of the key ingredients in the characterisation of stellar populations is the initial stellar mass function (IMF). In determining the link between the properties of the observed present-day mass function (MF) and the IMF, it is necessary to explore and account for the effects of GC dynamical evolution (mass loss and mass segregation) on the global and local (i.e. measured at various distances from a GC's centre) present-day MF and, in particular, determine how dynamical processes affect the evolution of the MF of stellar populations forming with different initial structural properties \citep[see e.g.][for simulations showing that the 2P forms more centrally concentrated than the 1P and \citealt{2018MNRAS.476.2731V}, Bruce et al., in prep. for studies of the evolution of the 1P and 2P MF]{2008MNRAS.391..825D,2010ApJ...724L..99B,2019MNRAS.489.3269C}. The study of the MF of multiple populations thus provides an avenue to address a number of fundamental questions concerning the formation and dynamical evolution of mPOPs in GCs.\looseness=-4

Very low mass stars also provide a particularly valuable laboratory for the chemical characterisation of mPOPs \citep[see e.g.][and references therein]{2024AN....34540018S}. Their cool and dense atmospheres are dominated by molecular opacity, which strongly affects their spectral energy distributions, especially in the near-infrared \citep{1995ApJ...445..433A,2002ApJ...568..335M}. As a consequence, variations in chemical composition can produce significant photometric and spectroscopic signatures in these stars. Furthermore, very low mass stars are fully convective and undergo minimal nuclear processing over their lifetimes, implying that their surface abundances largely reflect their initial chemical composition.\looseness=-4

In this work, we present a study of the MF of the different stellar populations in the GC 47\,Tucanae (NGC\,104, hereafter 47\,Tuc) covering a wide mass interval from $\sim 0.85$ down to $\sim 0.1\,M_\odot$. The paper is organised as follows: In Section\,\ref{Section2} we describe the dataset and the data-reduction procedures. In Section\,\ref{Section3} we present the MFs derived for the different stellar populations and discuss their theoretical interpretation. Finally, in Section\,\ref{Section4} we summarise our results and provide the main conclusions.\looseness=-4

\section{Observations and colour–magnitude diagram}\label{Section2}

The dataset used in this work consists of a combination of \textit{James Webb} Space Telescope (JWST) and \textit{Hubble} Space Telescope (HST) observations. The observations adopted in this paper are summarised in Table\,\ref{table:dataset}, while their field of view (FOV) is shown in Fig.\,\ref{FOV}. A brief description of the data reduction is provided in Appendix\,\ref{Appendix:A}.\looseness=-4

\begin{centering} 
\begin{figure}
 \includegraphics[width=\columnwidth]{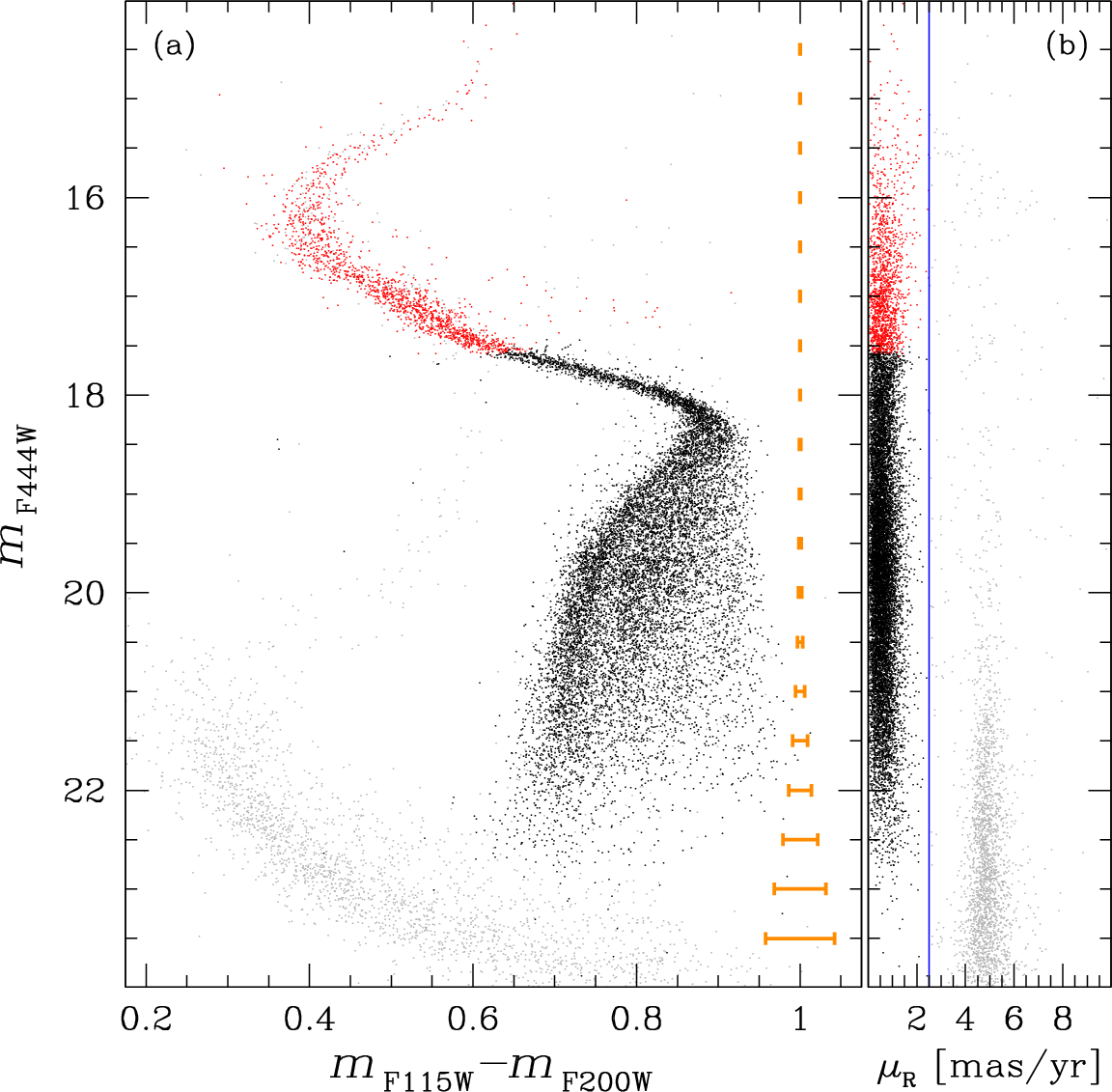}
 \caption{Colour-magnitude diagram of 47\,Tuc using JWST photometry. (a) $m_{\rm F444W}$ versus $m_{\rm F115W}-m_{\rm F200W}$ CMD. The orange error bars on the left side indicate the photometric uncertainties in colour derived from artificial stars as a function of the $m_{\rm F444W}$ magnitude. (b) $m_{\rm F444W}$ magnitude versus the one-dimensional PM ($\mu_{\rm R}$). The blue line separates cluster members from SMC and field stars. In both panels, sources that passed the PM selection are represented in black, while those that did not pass are shown in grey. Stars whose photometry is affected by saturation in at least one of the three filters are highlighted in red.\looseness=-4}  
 \label{cmd} 
\end{figure} 
\end{centering}

Panel\,(a) of Fig.\,\ref{cmd} shows the $m_{\rm F444W}$ versus $m_{\rm F115W}-m_{\rm F200W}$ CMD of our selected sample of stars. Photometry was corrected for the effects of differential reddening and spatial zero-point variations following the procedure introduced in \citet[][see also \citealt{2012A&A...540A..16M,2017ApJ...842....7B}]{2007AJ....133.1658S}. As shown in Fig.\,\ref{cmd}, the sequence remains narrow in the upper part of the CMD down to the knee ($m_{\rm F444W} \sim 18.25$), below which it broadens significantly. Stars below the MS knee have sufficiently low effective temperatures ($T_{\rm eff}\lesssim4000$ K) for H$_2$O molecules to form in their atmospheres. The F200W filter samples a spectral region strongly affected by H$_2$O absorption bands, while F115W is much less sensitive to oxygen-dependant molecular absorption \citep[][see also \citealt{2023A&A...679L..13C}]{2019A&A...629A..40S}. Since the strength of the H$_2$O absorption features depends on the amount of oxygen available, oxygen-rich stars experience stronger absorption in F200W than oxygen-poor stars. As a result, the colour $m_{\rm F115W}-m_{\rm F200W}$ becomes highly sensitive to oxygen abundance variations, producing the observed broadening. The broadened region shows a prominent concentration of stars on the blue side likely associated with oxygen-rich 1P stars and a more extended tail towards redder colours associated with oxygen-poor 2P stars with different degrees of oxygen depletion.\looseness=-4

To remove field stars and contaminants from the Small Magellanic Cloud (SMC), we measured the proper motions (PMs) of the stars from the displacement of stellar positions between the JWST GO-8816 and GO-2559 epochs, spanning a temporal baseline of $\sim2.4$\,yr. The resulting PMs are shown in panel\,(b) as the one-dimensional quantity $\mu_{\rm R}$ (defined as the quadrature sum of the two PM components) as a function of $m_{\rm F444W}$. Since the PMs are computed relative to the cluster’s bulk motion, cluster members are concentrated at low $\mu_{\rm R}$ values, while SMC stars form a distinct sequence around $\mu_{\rm R} \sim 5$\,mas\,yr$^{-1}$. We adopted a conservative selection criterion at $\mu_{\rm R} = 2.5$\,mas\,yr$^{-1}$ (vertical blue line). In both panels of Fig.\,\ref{cmd}, sources that satisfy this criterion are shown in black, while rejected objects are plotted in grey. Panel\,(a) further shows that SMC stars populate the lower-left region of the CMD. Saturated stars are also indicated in red in both panels.\looseness=-4

\section{The mass function of the different populations}\label{Section3}

\subsection{Observed mass function}\label{Section3.1}

We used the $m_{\rm F444W}$ versus $m_{\rm F115W}-m_{\rm F200W}$ CMD shown in Fig.\,\ref{cmd} to separate the different stellar populations in the lower MS (i.e. below the knee) of 47\,Tuc. To do that, we constructed a TpCD by combining the $m_{\rm F115W}-m_{\rm F200W}$ colour with $m_{\rm F150W2}-m_{\rm F322W2}$ (see panel\,(a) of Fig.\,\ref{cmd2}). The TpCD was built following the procedure adopted in similar previous studies \citep[see][and references therein]{2024A&A...688A.180S,2024A&A...689A..59S}. In brief, for each of the two CMDs, we defined fiducial lines enclosing the MS within a given magnitude interval. These fiducials were then used to verticalise the sequences, and the resulting normalised quantities were combined to construct the TpCD. For this analysis, we focused on the magnitude range $19 < m_{\rm F444W} < 22$ (corresponding to a mass range of $0.10 \lesssim M/M_{\odot} \lesssim 0.43$), where the separation between the two populations (1P and 2P) is most evident. The resulting TpCD is shown in panel\,(a) of Fig.\,\ref{tpcd}, with the corresponding density map in panel\,(b). These diagrams reveal a prominent clump of stars in the lower-right region of the TpCD around ($\Delta_{\rm F115W-F200W}$,$\Delta_{\rm F150W2-F322W2}$)$\sim$(0.8, 0.25) associated with the oxygen-rich 1P population and a less populated clump around $\sim$(0.6, 0.4) associated with the oxygen-poor 2P population, with an extended tail towards the upper-left region corresponding to increasingly extreme 2P stars. We defined boundaries by hand in the TpCD to isolate the two main populations. These boundaries were defined empirically from the stellar distribution in the TpCD and the corresponding density map to enclose the two main visible overdensities.\footnote{Additional substructures are visible within the 2P sample, likely indicating the presence of multiple 2P subpopulations; however, in this work, we focus only on the separation between 1P and 2P stars.} The resulting samples are shown in panel\,(c).\looseness=-4

To separate the two sequences in the upper part of the MS, we made use of the HST data. In particular, we constructed a TpCD by combining the $m_{\rm F438W}-m_{\rm F814W}$ colour with the pseudo-colour $C_{\rm F336W,F438W,F814W} = (m_{\rm F336W}-m_{\rm F438W})-(m_{\rm F438W}-m_{\rm F814W})$ -- see panels\,(b) and (c) of Fig.\,\ref{cmd2}. The pseudo-colour $C_{\rm F336W,F438W,F814W}$ is particularly sensitive to variations in light-element abundance and nitrogen abundance in particular, as the F336W filter includes strong NH molecular absorption bands, while the $m_{\rm F438W}-m_{\rm F814W}$ colour mainly traces temperature variations associated with differences in helium abundance \citep[see e.g.][]{2012ApJ...744...58M}. In the $m_{\rm F814W}$ versus $C_{\rm F336W,F438W,F814W}$ diagram shown in panel\,(b) of Fig.\,\ref{cmd2}, the sequence shows a mild splitting, with the bluer left-hand component corresponding to the N-poor 1P stars and the redder right-hand component corresponding to the N-enhanced 2P stars.\looseness=-4

The TpCD was built following the same procedure described above, focusing on the magnitude range $17.25 < m_{\rm F814W} < 19$ (corresponding to a mass range of $0.65 \lesssim M/M_{\odot} \lesssim 0.84$). The resulting TpCD is shown in panel\,(d), with the corresponding density map in panel\,(e). The diagram reveals a narrow clump centred around ($\Delta_{\rm C\,F336W,F438W,F814W}$,$\Delta_{\rm F438W-F814W}$)$\sim$(0.75, 0.25), associated with the N-poor He-normal 1P stars, and a more extended clump centred around $\sim$(0.3, 0.55), associated with the N-rich He-enhanced 2P stars. The two identified populations are highlighted in panel\,(f).\looseness=-4

We then derived the MF separately for the two identified stellar populations. The MFs were estimated using only stars in the common area covered by the JWST and HST observations spanning the radial range $1.64 \lesssim r/r_{\rm hl} \lesssim 2.62$ (see Fig.\,\ref{FOV}). Stellar magnitudes were converted into masses using the mass--luminosity relations presented by \citet{2024ApJ...961..139G,2025A&A...694A..68S}. In particular, we adopted the \texttt{G24B} mass--luminosity relation for the 1P stars and the \texttt{G24R} relation for the 2P stars. These two models differ primarily in their adopted oxygen abundances, with \texttt{G24B} representing an oxygen-rich composition ($[\mathrm{O/Fe}]=0.48$) and \texttt{G24R} representing an oxygen-poor one ($[\mathrm{O/Fe}]=0.03$). Masses were derived with the $m_{\rm F444W}$ magnitude and using the corresponding mass--luminosity relation.\looseness=-4

\begin{centering} 
\begin{figure}
 \includegraphics[width=\columnwidth]{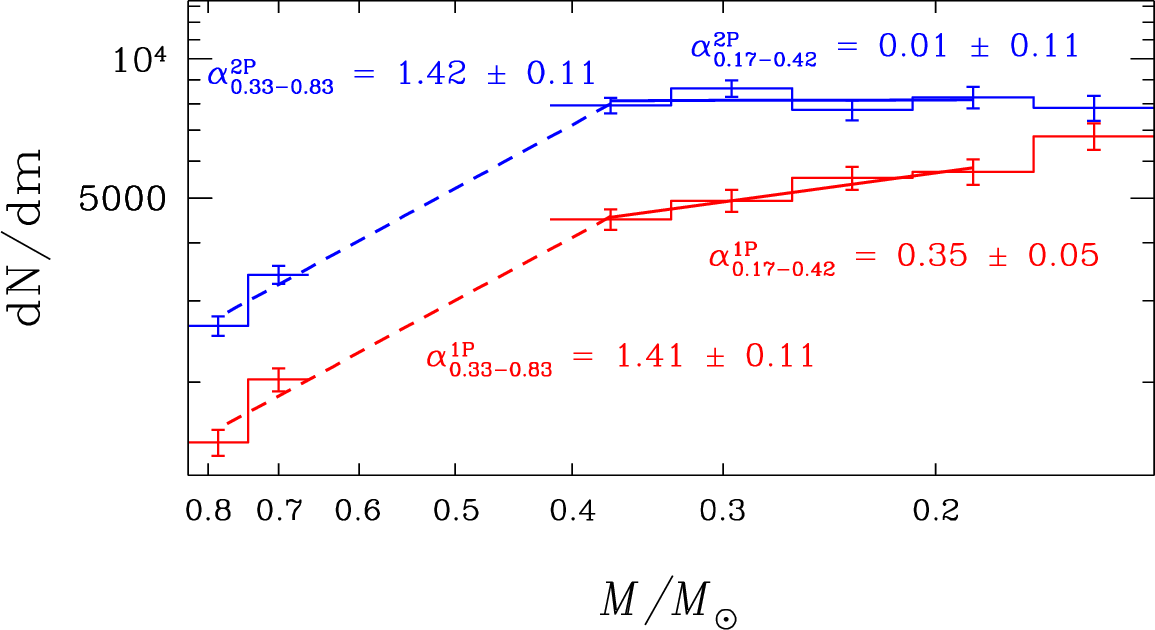}
 \caption{Mass functions of the two stellar populations in 47\,Tuc. The 1P and 2P populations are shown in red and blue, respectively. Histograms represent the completeness-corrected number of stars per unit mass. The solid lines show the best-fitting power-law relations in the low-mass regime, while the dashed lines represent the best-fitting power-law relations in the high-mass regime. The corresponding values of the MF slopes are labelled in the figure.\looseness=-4}
 \label{mf_pop} 
\end{figure} 
\end{centering}

Stars were divided into logarithmic mass bins, and for each bin we computed the number of stars belonging to each population, correcting for completeness. The resulting MFs are shown as histograms in Fig.\,\ref{mf_pop}, with the 1P and 2P populations represented in red and blue, respectively. To quantify the slopes of the MFs, we fitted the observed binned distributions with a power-law of the form $\xi(M) \propto M^{-\alpha}$, where $\xi(M)$ is the number of stars per unit mass and $\alpha$ is the slope. The fits were performed separately for the two populations over two distinct mass regimes. The first fit covers the lower-MS mass range $0.17 \lesssim M/M_{\odot} \lesssim 0.42$. The second fit combines the highest-mass JWST bin with the two HST bins, spanning the mass interval $0.33 \lesssim M/M_{\odot} \lesssim 0.83$. For the low-mass regime, the fit excludes the lowest-mass bin, where the MF shows the break observed in \citet{2025A&A...701A.169S}. The best-fitting relations are shown in Fig.\,\ref{mf_pop}, respectively, and the corresponding values of $\alpha$ are reported in the figure.\looseness=-4

As can be seen from Fig.\,\ref{mf_pop}, in the low-mass regime, the MF of the 1P population is significantly steeper ($\alpha^{\rm 1P}_{0.17-0.42}=0.35\pm0.05$) than that of the 2P population, which is nearly flat ($\alpha^{\rm 2P}_{0.17-0.42}=0.01\pm0.11$). Over the mass interval $0.33 \lesssim M/M_{\odot} \lesssim 0.83$, the two populations show very similar MF slopes, with $\alpha^{\rm 1P}_{0.33-0.83}=1.41\pm0.11$ and $\alpha^{\rm 2P}_{0.33-0.83}=1.42\pm0.11$. In the next subsection, we compare the obtained slopes with predictions from theoretical simulations.\looseness=-4

\subsection{Theoretical interpretation of the mass functions slopes}\label{Section3.2}

To investigate the origin of the flatter low-mass MF observed in the 2P relative to the 1P, we ran a Monte Carlo simulation using the \texttt{MOCCA} code \citep[][]{Giersz1998, Hypki+2013}, which follows the long-term dynamical evolution of GCs. Rather than assuming the conventionally adopted universal \citet{2001MNRAS.322..231K} IMF, we initialised our model with a three-parameter IMF that differs between the 1P and 2P. The choice of a three-parameter bottom-light IMF is motivated by previous studies suggesting that massive GCs such as 47\,Tuc are better described by a deficiency of low-mass stars relative to the standard \citet{2001MNRAS.322..231K} IMF \citep[][Dalessandro et al., in prep.]{Giersz+2011, Webb+2017, Cadelano+2020, Baumgardt+2023}, while the introduction of a population-specific IMF is motivated directly by the observational results of the present work. We emphasise that this simulation is intended to illustrate the dynamical evolution of the stellar MF in a cluster of comparable age and mass to 47\,Tuc at 12\,Gyr and does not constitute a direct model of 47\,Tuc itself.\looseness=-4

Specifically, the IMF adopted in our model assumes 1P (2P) slopes of $0.5$ ($0.05$) in the mass range $0.1$--$0.4\,M_\odot$, slopes of $1.65$ ($1.65$) in the mass range $0.4$--$1.0\,M_\odot$, and slopes of $2.3$ ($2.3$) in the mass range $1.0$--$100\,M_\odot$. This is an extension of the IMF proposed by \citet[][]{Baumgardt+2023}, modified in the low-mass regime to incorporate a flatter slope for the 2P and a steeper slope for the 1P. The adoption of this difference between the slopes of the 1P and 2P IMF is necessary since in dynamical models starting with the same 1P and 2P IMF \citep[][Bruce et al., in prep.]{2018MNRAS.476.2731V}, the slope of the 2P present-day MF remains similar or (in the outer regions) becomes steeper than the 1P MF, which is in contrast to our observational finding in the low-mass range. To illustrate this point, we also present results from a simulation that starts with the same IMF for both the 1P and 2P stars, adopting in this case the IMF suggested by \citet{Baumgardt+2023}, which has slopes equal to 0.3, 1.65, and 2.3 in the three mass ranges indicated above. Both simulations start with $N=2.8\times10^6$ objects (where an object is a single star or a binary star) consisting of $2.1\times10^6$ 1P objects and $0.7\times10^6$ 2P objects (the fraction of 2P objects remains approximately constant over the cluster's evolution), with a 5\% binary fraction and an initial half-mass radius equal to 1 pc. The initial spatial distribution of the 2P (1P) system follows that of a \citet{King1966} model with a central dimensionless potential $W_0=7$ ($W_0=5$) and an initial ratio of the 2P to the 1P half-mass radii equal to 0.1. We show the radial profile of the slope of the mass function of our model at 12 Gyr, at which point the cluster has reached a dynamical age of $t/t_{rh}(t)\sim4$ and a stellar mass of $M_\star\sim10^6M_\odot$, is broadly consistent with the values measured for 47\,Tuc.\looseness=-4 

\begin{centering} 
\begin{figure}
 \includegraphics[width=\columnwidth]{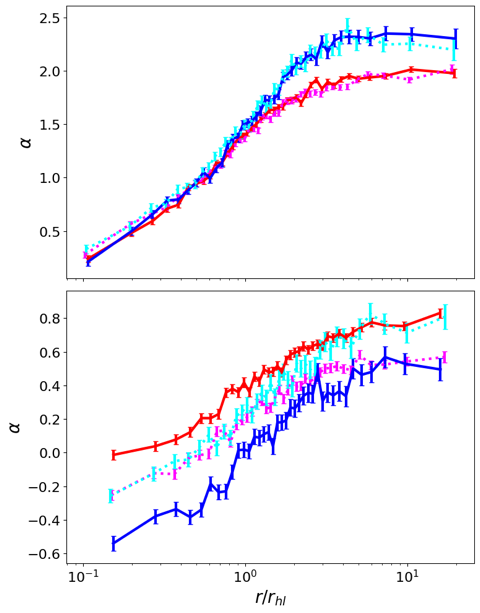}
 \caption{Slope of the stellar MF ($\alpha$) as a function of projected distance from the cluster centre normalised to the projected half-light radius for 1P (red) and 2P (blue) at 12 Gyr in the mass range $0.33$--$0.83\,M_\odot$ (top panel) and $0.17$--$0.42\,M_\odot$ (bottom panel). Data from the simulation starting with the same 1P and 2P IMF for the entire mass range are shown in magenta (1P) and cyan (2P).\looseness=-4}
 \label{simulation} 
\end{figure} 
\end{centering}

The top panel of Figure~\ref{simulation} shows the radial variation of the MF slope at 12 Gyr as a function of projected distance from the cluster centre, normalised to $r_{\rm hl}$, for the upper mass range $0.33$--$0.83\,M_\odot$. In this regime, the IMF does not differ between the two populations (the IMF differs only in the mass range $0.1$--$0.4\,M_\odot$), and the profiles evolve as expected. Both populations develop a radial gradient that is flatter in the inner regions, dominated by inward segregation of massive stars, and steeper in the outer regions, dominated by outward migration of lower-mass stars. A notable difference between the populations does emerge in the outer regions. While the 1P slope there remains closer to its IMF value, the 2P slope is significantly steeper, owing to the preferential outward migration of low-mass 2P stars from a central distribution to regions initially devoid of them (see also Bruce et al., in prep.). In an extended radial region of the model, the values of $\alpha$ in the 1P and 2P are similar to each other, as found in our observational analysis. In our model the specific observed values of $\alpha$ are found at smaller clustercentric distances, but as pointed out above, our simulation is not aimed at providing a detailed model for 47 Tuc. The bottom panel of Figure~\ref{simulation} shows the radial variation of the MF slope at 12 Gyr in the lower mass range $0.17$--$0.42\,M_\odot$, the regime in which the 2P IMF is intrinsically flatter than that of 1P. The qualitative shape of the radial profiles is similar for both populations (i.e. flatter in the centre and steeper in the outskirts) but with an important distinction: Across all radii, the 2P slope remains systematically flatter than that of the 1P slope. This offset is a direct imprint of the population-specific IMF, and it cannot be reproduced by dynamical evolution alone starting from a universal cluster IMF. The difference between the 1P and 2P MF predicted by this simulation is consistent with that found in our observational analysis, which is difficult to reconcile without invoking a flatter 2P IMF in the lower-mass regime. Indeed, as shown in the bottom panel of Figure~\ref{simulation}, in the model starting with the same 1P and 2P IMF for the entire mass range, the difference between the 1P and 2P slopes is opposite to that found in observations.\looseness=-4

\section{Summary and conclusion}\label{Section4}
In this work, we have used a combination of JWST and HST observations covering the same region of 47\,Tuc spanning the radial range $1.64 \lesssim r/r_{\rm hl} \lesssim 2.62$ to separate the two main stellar populations over a wide stellar-mass interval, extending from $\sim0.85\,M_{\odot}$ down to $\sim0.1\,M_{\odot}$. While the separation of multiple populations in 47\,Tuc below the MS knee was previously demonstrated using a combination of JWST and HST optical/near-UV photometry \citep[see e.g.][]{2023MNRAS.522.2429M}, here we showed that the two populations can be efficiently separated below the MS knee using JWST data alone, without the need for complementary HST observations.\looseness=-4

We derived the MFs of the two identified populations across the entire mass extent. In the upper mass range of this work ($0.33$--$0.83\,M_{\odot}$), the slopes of the 1P and 2P present-day MFs are similar to each other and consistent with what has been found in the dynamical model presented in this paper (see Fig.\,\ref{simulation}). In the low-mass regime ($0.17$--$0.42\,M_{\odot}$), on the other hand, the MF of the 1P population is significantly steeper than that of the 2P population, which appears nearly flat (see Fig.\,\ref{mf_pop}). This difference is contrary to what is expected from models accounting for the effects of dynamical evolution of multiple populations starting with the same IMF for the 1P and the 2P. The results of the Monte Carlo dynamical model presented in this paper instead show that this observational finding can be explained with dynamical models starting with a 2P IMF flatter (in the low-mass regime) than the 1P IMF (see Fig.\,\ref{simulation}). Our findings therefore indicate that the two main multiple populations in 47\,Tuc formed with different IMFs. The results presented in this work thus provide a new fundamental constraint on the star formation of multiple populations in GCs. Future studies extending this investigation to a broader range of radial clustercentric distances will be important to further exploring the differences revealed in our investigation and providing new constraints for models of formation and dynamical evolution of multiple populations.\looseness=-1

\begin{acknowledgements}
EV acknowledges support from NSF grant AST-2009193.
\end{acknowledgements}
\looseness=-4
\bibliographystyle{aa}
\bibliography{main.bib}

\begin{appendix}
\onecolumn
\section{Dataset, reduction, and field coverage}\label{Appendix:A}

The data reduction follows the procedures adopted in previous works based on both HST \citep[see][for details]{2021MNRAS.505.3549S} and JWST \citep[see][for details]{2022MNRAS.517..484N,2023MNRAS.525.2585N,2023AN....34430006G} observations. In brief, we made use of calibrated, un-resampled exposures (\texttt{\_flc} and \texttt{\_cal} fits files for HST and JWST, respectively, with the latter obtained from level-1b \texttt{\_uncal} exposures using a development version of the JWST pipeline).

Stellar photometry and astrometry were obtained through a two-step procedure consisting of a first and a second pass. Photometry was then calibrated onto the VEGA-mag system, and the astrometry was anchored to the \textit{Gaia} Data Release 3 \citep[DR3][]{2016A&A...595A...1G,2023A&A...674A...1G} absolute reference frame.

We selected well-measured stars using diagnostic parameters such as the quality-of-PSF fit parameter (\texttt{QFIT}), the excess or deficiency of the source’s flux \citep[\texttt{RADXS}, see][]{2008ApJ...678.1279B,2009ApJ...697..965B}, and the local sky noise (\texttt{rmsSKY}).

The completeness of our catalogue was estimated using artificial stars (ASs), following procedures adopted in similar works \citep{2025A&A...694A..68S,2025A&A...701A.169S}. In brief, ASs were generated along fiducial lines tracing the two populations in multiple CMDs to ensure coverage of all filters in our dataset, and were injected, detected, and measured using the same procedures adopted for real stars. An AS was considered recovered if its output position and magnitude were consistent with the input values and if it satisfied the same photometric-quality and PM selection criteria adopted for observed sources. Completeness was estimated as the fraction of recovered ASs as a function of magnitude.

\begin{table*}[h!]
\centering
\begin{threeparttable}
\caption{Summary of the dataset used in this work.}
\label{table:dataset}

\begin{tabular}{l c c c c}
\hline
\hline
Filter & Instrument & $N \times$ Exp. time & Date & Programme ID\\
\hline

\multicolumn{5}{c}{HST}\\
\hline

F336W  & WFC3/UVIS & $2\times300$\,s & Sep. 4, 2015 -- Mar. 11, 2016 & 14021$^{a}$\\
F438W  & WFC3/UVIS & $1\times35 + 9\times350$\,s & Jul. 17, 2009 & 11452$^{b}$\\
F438W  & WFC3/UVIS & $2\times800$\,s & Aug. 3, 2010 & 11903$^{c}$\\
F606W  & ACS/WFC    & 4$\times$1\,s+4$\times$10\,s+4$\times$100\,s+117$\times$1113\,s–1498\,s & Jan. 15 -- Oct. 1, 2010 & 11677$^{d}$\\
F814W  & ACS/WFC    & 4$\times$1\,s+4$\times$10\,s+4$\times$100\,s+125$\times$1031\,s-1484\,s & Jan. 15 -- Oct. 1, 2010 & 11677$^{d}$\\

\hline

\multicolumn{5}{c}{JWST}\\
\hline

F070W  & NIRCam & 9$\times$236\,s & Oct. 22, 2025 & 8816$^{e}$\\
F115W  & NIRCam & 9$\times$236\,s & Oct. 22, 2025 & 8816$^{e}$\\
F150W2 & NIRCam & 6$\times$945\,s & May 24--25, 2023 & 2559$^{f}$\\
F200W  & NIRCam & 9$\times$236\,s & Oct. 22, 2025 & 8816$^{e}$\\
F277W  & NIRCam & 9$\times$236\,s & Oct. 22, 2025 & 8816$^{e}$\\
F322W2 & NIRCam & 6$\times$945\,s & May 24--25, 2023 & 2559$^{f}$\\
F356W  & NIRCam & 9$\times$236\,s & Oct. 22, 2025 & 8816$^{e}$\\
F444W  & NIRCam & 9$\times$236\,s & Oct. 22, 2025 & 8816$^{e}$\\

\hline
\end{tabular}

\begin{tablenotes}
\footnotesize
\item[] \textit{Principal Investigator:}
(a)~Deustua;
(b)~Quijano;
(c)~Kalirai;
(d)~Richer;
(e)~Rest;
(f)~Caiazzo.
\end{tablenotes}

\end{threeparttable}
\end{table*}

\begin{centering} 
\begin{figure}[h!]
 \centering
 \includegraphics[width=.4\columnwidth]{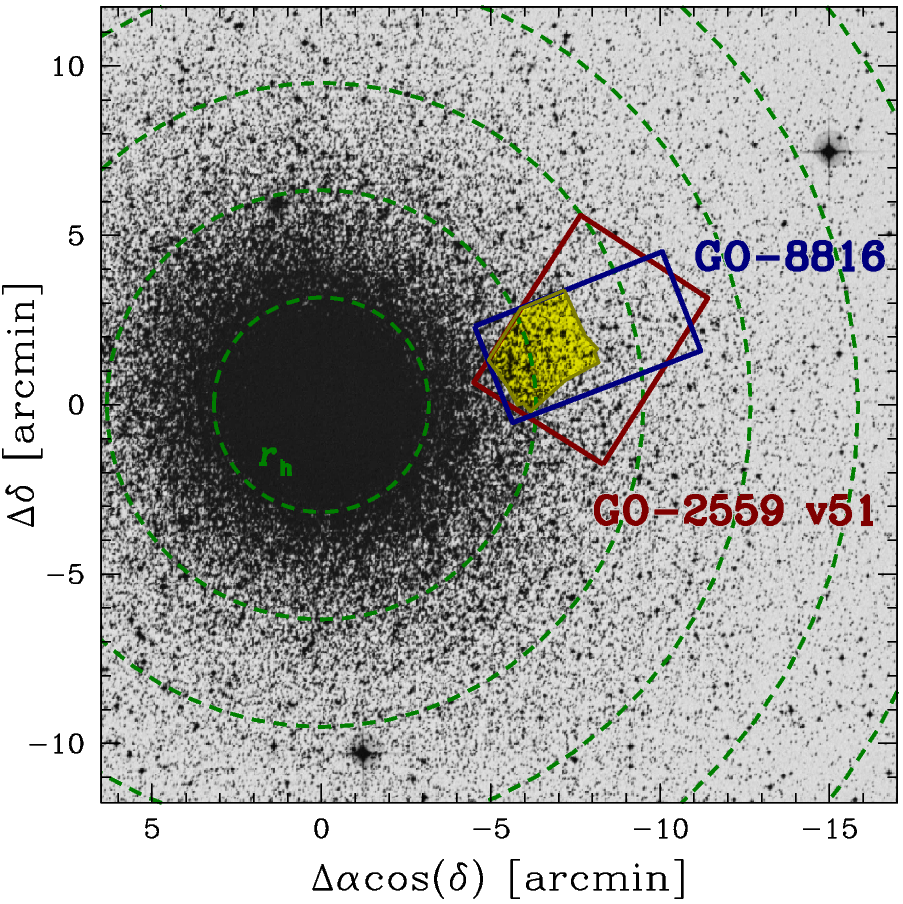}
 \caption{Footprints of the JWST observations used in this work overlaid on a Digital Sky Survey (DSS) image of 47\,Tuc. The JWST GO-8816 and the second epoch of the GO-2559 observations (visit v51) NIRCam fields are shown in blue and red, respectively, while the common area between the JWST and HST observations is highlighted in yellow. Axes are given in arcminutes from the cluster centre. The green dashed circles mark the half-light radius of the cluster \citep[$r_{\rm hl}=3\farcm17$;][]{2010arXiv1012.3224H} and its multiples.} 
 \label{FOV} 
\end{figure} 
\end{centering}

\onecolumn
\section{Additional colour--magnitude diagrams and population separation}\label{Appendix:B}

\begin{centering} 
\begin{figure*}[h!]
 \includegraphics[width=.95\textwidth]{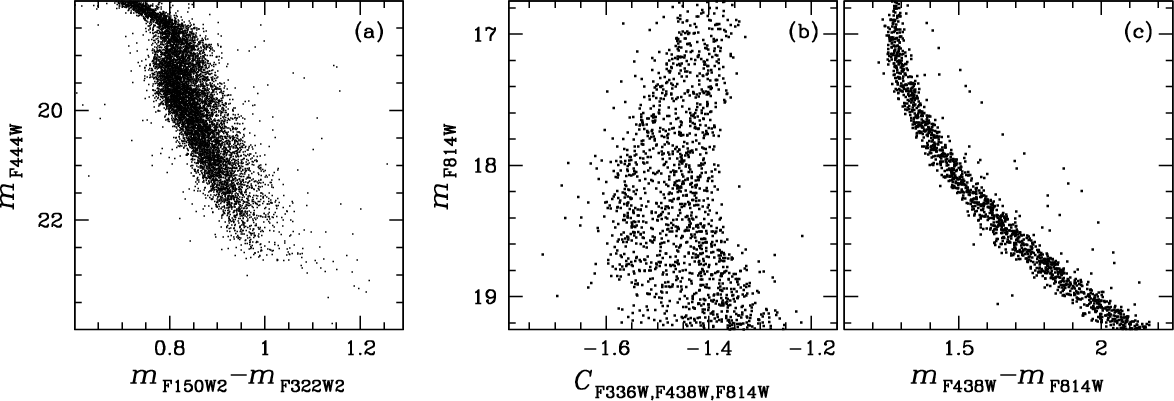}
 \caption{Colour--magnitude diagrams of 47\,Tuc. (a) $m_{\rm F444W}$ versus $m_{\rm F150W2}-m_{\rm F322W2}$ CMD in the lower-MS regime, where the sequence broadens below the MS knee as a consequence of oxygen abundance variations. (b)-(c) $m_{\rm F814W}$ versus $C_{\rm F336W,F438W,F814W}$ and $m_{\rm F814W}$ versus $m_{\rm F438W}-m_{\rm F814W}$ CMDs.}  
 \label{cmd2} 
\end{figure*} 
\end{centering}

\begin{centering} 
\begin{figure*}[h!]
 \includegraphics[width=.95\textwidth]{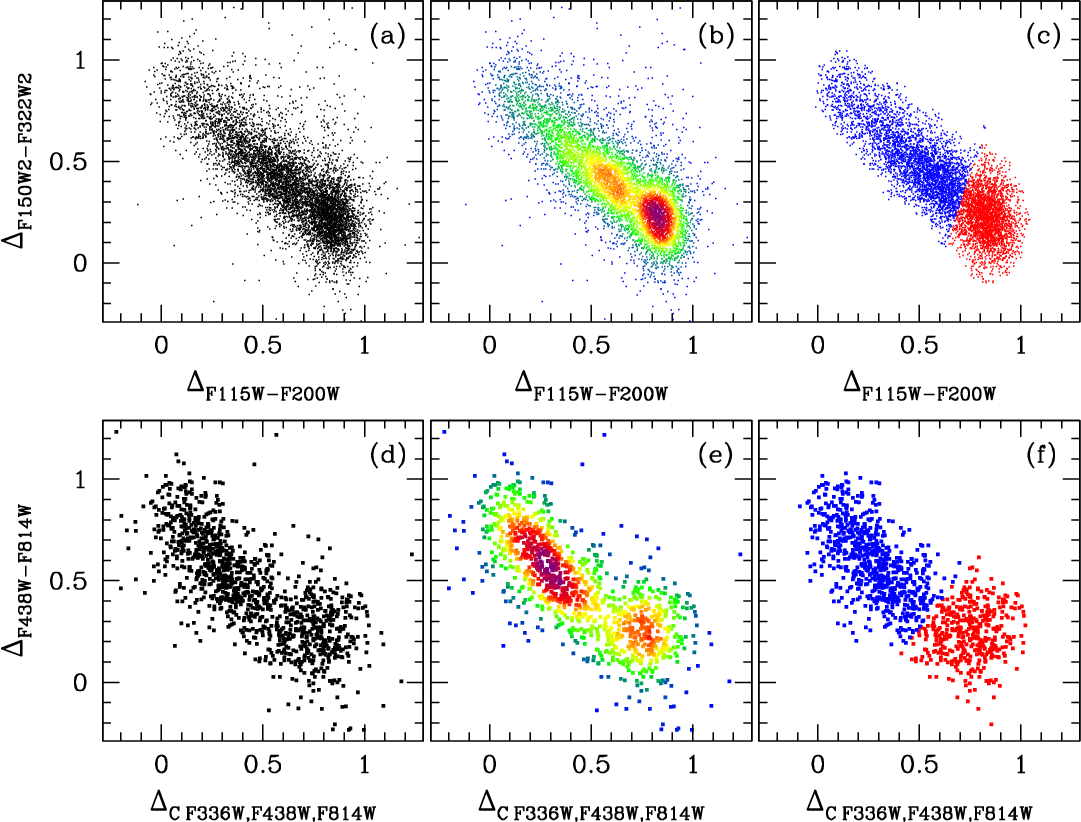}
 \caption{Two-pseudo-colour diagrams used to separate the mPOPs in 47\,Tuc. Panels\,(a) and (d) show the TpCDs for the lower ($0.10 \lesssim M/M_{\odot} \lesssim 0.43$) and upper ($0.65 \lesssim M/M_{\odot} \lesssim 0.84$) MS, respectively. Panels\,(b) and (e) present the corresponding density maps. The colour goes from blue (lowest density) to green (average density) to red (highest density). Panels\,(c) and (f) show the selected samples of stars, with the two populations, 1P and 2P, highlighted in red and blue, respectively. The association between the populations identified in the two mass intervals follows the association adopted in similar studies of 47\,Tuc \citep[see e.g.][and references therein]{2025A&A...698A.247M}. The top panels are based on JWST photometry, while the bottom panels are based on HST photometry. $\Delta$ denotes the verticalised colour or pseudo-colour.}
 \label{tpcd} 
\end{figure*} 
\end{centering}

\end{appendix}

\end{document}